\documentclass[usegraphicx,usenatbib,onecolumn]{mn2e}
\usepackage{amssymb}
\usepackage{bm}

\begin{document}
\title{Detectability of the 21 cm-CMB cross-correlation from the
  EoR} 

\author[Tashiro, H. et al.]
{Hiroyuki Tashiro$^1$, Nabila Aghanim$^1$, Mathieu Langer$^1$, 
\newauthor Marian Douspis$^1$, Saleem Zaroubi$^2$, and Vibor Jelic$^2$ \\
  $^1$ Institut d'Astrophysique Spatiale (IAS), B\^atiment 121, 
  Universit\'e Paris-Sud XI and CNRS, F-91405, Orsay (France);\\
$^2$ Kapteyn Astronomical Institute, University of Groningen, Postbus 800,
NL-9700AV, Groningen, (The Netherlands)
}

\date{\today}
\maketitle

\begin{abstract}
The 21-cm line fluctuations and the cosmic microwave background (CMB) 
are powerful probes of the epoch of reionisation (EoR) of the universe. We study the
potential of the cross-correlation between 21-cm line fluctuations and
CMB anisotropy to obtain further constraints on the
reionisation history. We compute analytically
the 21-cm cross-correlation with the CMB temperature anisotropy 
and polarisation, and we
calculate the signal-to-noise (SN) ratio 
for its detection with $\it{ Planck}$ together with
LOFAR, MWA and SKA.
We find, on the one hand, that the 21-cm cross-correlation signal with
CMB temperature from the instant reionisation can be detected with an
SN ratio of $\sim 2$ for LOFAR and $\sim 10$ for SKA.  On the other
hand, we confirm that the detection of the 21-cm cross-correlation
with CMB polarisation is practically infeasible.

\end{abstract}

\begin{keywords}
cosmology: theory -- cosmic microwave background -- large-scale structure of universe

\end{keywords}

\maketitle

\section{Introduction}

The measurement of 21~cm line of neutral hydrogen from high redshifts
is eagerly awaited as a probe of the Epoch of Reionisation (EoR). 
During the EoR, the first collapsed objects
heat and ionise the intergalactic medium (IGM).
Therefore, the epoch and the process of reionisation are
tightly related to the
evolution of cosmological structure and the formation of the
first objects \citep{barkana-loeb-2001,ciardi-ferrara-2005,
fan-carilli-2006}.
The 21~cm fluctuations are sensitive to the density, temperature,
and ionised fraction of IGM.
Studying the 21-cm tomography 
tells us about
the physics of IGM gas and structure formation during the EoR
\citep{madau-meiksin-rees-1997,
tozzi-madau-2000,ciadri-madau-2003,furlanetto-zaldarriaga-2004},
and several 21-cm experiments are recently designed and built (e.g.
MWA\footnote{http://web.haystack.mit.edu/array/MWA},
LOFAR\footnote{http://www.lofar.org},
SKA\footnote{http://www.skatelescope.org}).


The 21~cm cross-correlation with other complementary probes is
expected to provide additional information other than their respective
auto-correlations.  Besides, the cross-correlation has an advantage
for observations of 21~cm fluctuations whose signal is weak, because
it suffers from foregrounds and systematic effects less than the
auto-correlations.  The cross-correlation between the 21~cm line and
the cosmic microwave background (CMB) has been studied by many authors.  On large scales ($\ell \sim
100$), the 21-cm fluctuations cross-correlate with the CMB Doppler
temperature anisotropies which are due to the motions of ionised
baryons \citep{alvarez-komatsu,adshead-furlanetto-2007}.  Because the
maximum amplitude of the cross-correlation is reached at the redshift
when the ionised fraction is one half, it is sensitive to the EoR.  On
small scales ($\ell >1000$), cross-correlation between the 21~cm
fluctuations and CMB temperature anisotropies from reionisation
bubbles arises
\citep{salvaterra-ciardi,cooray-2004,slosar-cooray-2007,jelic-zaroubi-2009}.
\citet{salvaterra-ciardi} showed that these two signals are
anti-correlated on the scale corresponding to the typical size of an
ionised bubble.  \citet{tashiro-polar-2008} studied the 21~cm
cross-correlation with CMB $E$-mode polarisation on large scales.
They have shown that the peak of the cross-correlation spectrum
reaches its maximum value when the average ionised fraction of the
universe is about half as shown in the case of the 21~cm
cross-correlation with the CMB Doppler temperature, and there is a
damping that depends on the duration of reionisation.  The
cross-correlation between the 21~cm fluctuations and high redshift
galaxy distribution has also the potential to probe the EoR
\citep{Wyithe-Loeb-2007,furlanetto-lidz-2007,Lidz-Zahn-2009}.  On
large scales, the 21-cm and galaxy distributions are anti-correlated,
while on scales smaller than the typical size of an ionised bubble,
these fields become roughly uncorrelated.  Therefore, the
cross-correlation between the 21~cm fluctuations and high redshift
galaxy distributions provides access to the evolution of the typical
scale of the ionised bubble.

In this paper, we investigate the detectability of the
cross-correlation between the 21~cm fluctuations and CMB by performing
a signal-to-noise (SN) analysis.  Particularly, we focus on the
cross-correlation on large scales and we discuss the detectability of
the signals and the sensitivity to the reionisation properties by
Planck and LOFAR which will release useful data for the
cross-correlation in the near future.  This article is organised as
follows.  In Sec.~II, we give a short description of the SN analysis.
In Sec.~III, we provide analytic form of 
the cross-correlation between the 21~cm fluctuations 
and the CMB anisotropy which include both CMB temperature anisotropy and CMB
$E$-mode polarisation.
In Sec.~IV, we show 
the angular power spectrum of the cross-correlation.
In Sec.~V, we evaluate the the SN ratio of the cross-correlation and
discuss the detectability by upcoming observations.
Sec.~VI is devoted to the conclusions.  Throughout the paper, we use
WMAP 5-year values for the cosmological parameters, i.e. $h=0.73 \ (H_0=h
\times 100 {\rm km/s / Mpc})$, $T_0 = 2.725$K, $h^2 \Omega _{\rm b}
=0.0223$ and $h^2 \Omega_{\rm m} =0.128$ \citep{wmap5-cosmo} for a
flat cosmology.

\section{Signal-to-Noise analysis}\label{sec:errorestimate}

In order to investigate the detection level of the 
signals, a useful tool is the signal-to-noise (SN) analysis. 
The SN analysis not only can give the prospective detection level for the observations 
but also allows us to compute the optimal observational properties 
for an arbitrary detection level. 

In order to evaluate SN ratio, first, we must estimate the error of
the power spectrum of the cross-correlation.  For simplicity, we
assume that CMB, 21-cm fluctuations and instrumental noise are
Gaussian and the foregrounds and noise of 21-cm fluctuations and CMB
anisotropy are not correlated.  Under these assumption, the error of
the power spectrum of the cross-correlation can be rewritten as
\citep{knox-1995}
\begin{equation}
\Delta C_\ell
^2 = 
\frac{1}{(2\ell+1)f_{\rm sky}\Delta \ell}
\left[(C_l^{\rm 21-\alpha})^2+(C_\ell^{\alpha}+ N_\ell^{\alpha}) 
(C_\ell^{21}+ N_\ell^{21})\right],
\label{eq:error}
\end{equation}
where 
the superscript $21$ stands for 21-cm fluctuations 
and the superscript $\alpha$ stands for $D$, the CMB Doppler anisotropy,
or $E$, the $E$-mode polarisation,
and $C_\ell$ and $N_\ell$ are the signal from the EoR and 
the noise power spectrum, respectively. 
In Eq.~(\ref{eq:error}),
$\Delta l$ is the size of bins within which the power spectrum
data are averaged over $l-\Delta l/2<l<l+\Delta l/2$, and
$f_{\rm sky}$ is the sky fraction
common to the two cross-correlated signals.
In this paper, we consider {\it Planck} as CMB observation, which is almost full-sky.
Therefore $f_{\rm sky}$ corresponds to the sky fraction of  21~cm observations 
which is the order of a few percents at most.

From Eq.~(\ref{eq:error}),
we can obtain the total SN ratio for the 21-cm cross-correlation as
\begin{equation}
\left( {S \over N} \right) ^2 =
f_{\rm sky} \sum_{\ell = \ell_{\rm min}} ^{\ell_{\rm max}} (2 \ell +1)
{| C_\ell ^{21-\alpha}| ^2  \over | C_\ell ^{21-\alpha} |^2
+(C_\ell ^{21} +N_\ell ^{21} ) (C_\ell ^{\alpha} +N_\ell ^{\alpha})}.
\label{eq:SNratio}
\end{equation}
In the next section, we discuss the cross-correlation signal from reionisation
and we explicit the noise power spectrum in Sec.~\ref{sec:noise}.

\section{Formalism of the cross-correlation}\label{sec:cross}

The angular power spectrum of the cross-correlation between
21~cm fluctuations and CMB has been obtained by \citet{alvarez-komatsu}
and \citet{tashiro-polar-2008}.
Here we recall the analytic form of the cross-correlation with 
CMB Doppler temperature anisotropy and $E$-mode polarisation
and give a short description for our reionisation model.

\subsection{21~cm line fluctuations}

The observed brightness temperature of the 21~cm lines in a direction
$\hat {\bm n}$ and at a frequency $\nu$ is given as in
\cite{madau-meiksin-rees-1997} by
\begin{equation}
T_{21} (\hat {\bm n};\nu) = \frac{\tau_{21}}{(1+z_{\rm obs})}
(T_{\rm s} -T_{\rm CMB})(\eta_{\rm obs}, \hat {\bm n} (\eta_0-\eta_{\rm
obs})),
\label{eq:21cmline}
\end{equation}
where $T_{\rm CMB}$ is the CMB temperature and  $T_{\rm s}$
is the spin temperature given by the ratio of the number
density of hydrogen in the excited state to that of hydrogen in the
ground state.  The conformal time $\eta_{\rm obs}$ is associated with the redshift $z_{\rm
obs}$ and  $\nu = \nu_{21}/(1+z_{\rm obs})$ with $\nu_{21}$ being
the frequency corresponding to the 21~cm wavelength. 
The optical depth for the 21~cm line absorption $\tau_{21}$ is
\begin{equation}
\tau_{21} 
= {3 c^3 \hbar A_{10} x_{\rm H} n_{\rm H} \over 16 k \nu_{21} ^2 T_{\rm s} H(z)}
\label{eq:tau21}
\end{equation}
where $n_{\rm H}$ is the hydrogen number density 
and $x_{\rm H}$ is the fraction of neutral hydrogen, which is written as a
function of the ionised fraction $x_e = 1- x_{\rm H}$.

According to Eqs. (\ref{eq:21cmline}) and (\ref{eq:tau21}), the observed
brightness temperature of the 21~cm lines will reflect 
baryon density fluctuations, $\delta _{\rm b} \equiv (\rho_b - \bar \rho_b)/\bar \rho_b$, and 
fluctuations of the neutral hydrogen fraction, 
$\delta _{H} \equiv (x_H - \bar x_H)/\bar x_H$,
where $\rho_b$ is the baryon density and 
the symbols with a $^{-}$ represent the background values. 
We can rewrite
Eq.~(\ref{eq:21cmline}) in the linear approximation
\begin{equation}
T_{21} (\hat {\bm n} ;\nu) = 
[1- \bar x_e(1+ \delta_{ x})](1+\delta_{\rm b}) T_{0}
\approx [(1- \bar x_e)(1+\delta_{\rm b}) + \bar x_e \delta_{ x}] T_{0},
\label{eq:21cmlinefl}
\end{equation}
where $\bar x_e$ and $\delta_{ x}$ are the average 
and the fluctuations of the ionised fraction, respectively,
which are $\bar x_e = 1- \bar x_H $ and $\delta_{ x}=-\delta_{ H}$ 
in the linear approximation,
and  $T_0$ is a normalisation temperature factor given by 
\begin{equation}
T_0 =23  \left({\Omega_{\rm b} h^2 \over 0.02} \right)
\left[ \left({0.15\over \Omega_{\rm m} h^2} \right)  
\left( {1+z_{\rm
      obs} \over 10} \right) \right] ^{1/2} \left({T_{\rm s}-T_{\rm cmb} \over T_{\rm s}} \right)~{\rm mK}. 
\end{equation}
The spin temperature is determined by three couplings with CMB, gas and Ly-$\alpha$ photons.
Before the reionisation, $T_{\rm s}$ is set by the balance between 
the couplings with CMB and gas.
Then, after gas is heated by stars and QSOs and the reionisation starts 
$T_{\rm s}$ becomes much larger than the CMB temperature 
mainly by the Ly-$\alpha$ coupling 
\citep{ciadri-madau-2003}.
In this paper, since we focus on 21~cm signals from the EoR,
we assume $T_{\rm s} \gg T_{\rm cmb}$ in order to obtain $T_0$. 


The 21~cm line fluctuation map at a frequency $\nu$ 
can be described by
\begin{equation}
\delta T_{21}( \hat {\bm n};\nu ) = T_{0} \sum_\ell
\int \frac{dk^3}{(2 \pi)^3}\sqrt{4 \pi (2 \ell +1)}  
\left [(1- \bar x_e) (1+ F \mu^2)\delta_{\rm b} - \bar x_e \delta_{
    x}
\right ]
j _\ell(k (\eta_0-\eta_{\rm obs})) Y_\ell ^0( \hat{\bm n}),
\label{eq:21cm-map}
\end{equation}
where we take the Fourier expansion of $\delta_{\rm b}$ and $\delta_{x}$
with Rayleigh's formula.
We also introduced the factor $(1+ F
\mu^2)$ to account for the enhancement of the fluctuation amplitude
due to the redshift distortion (Kaiser effect) on the 21~cm line
fluctuations, $\mu = \hat {\bm k} \cdot \hat {\bm n}$ and $F=d \ln g /
d \ln a$ with $g(a)$ the linear growth factor of baryon fluctuations
\citep{bharadwaj-ali-2004}.

\subsection{CMB anisotropy}

As reionisation proceeds, the coupling of CMB photons and free electrons
by Thomson scattering becomes strong again.
As a result, Thomson scattering during reionisation produces secondary
CMB temperature anisotropy and polarisation. 

In the CMB temperature,
the main generation mechanisms at the EoR
are the Doppler effect for first order anisotropic fluctuations and 
the kinetic Sunyaev-Zel'dovich effect
for the second order.
While the former is dominant on large scales ($\ell < 1000$),
the latter dominates on small scales ($\ell > 1000$).
In the following, we focus on the computation of the cross-correlation
power spectrum on large scales ($\ell \sim 100$). We therefore 
consider only the Doppler anisotropy and
neglect the kinetic Sunyaev-Zel'dovich effect, 
although, by making this hypothesis, we underestimate 
the CMB temperature anisotropy generated during reionisation at $\ell \sim 1000$.

The Doppler anisotropy of the CMB temperature produced during the EoR 
is given by
$
T_{\rm D}({\bf\hat{n}})= -T_{\rm cmb}
\int_0^{\eta_0} d\eta \dot{\tau}e^{-\tau}{\bf\hat{n}}\cdot {\bf
  v}_{\rm b}({\bf\hat{n}},\eta)
$
where $\dot \tau$
is the differential optical depth for Thomson scattering $\tau(\eta)$
in conformal time $\dot \tau = n_e \sigma_{\rm T}a$ with the electron
number density $n_e$, the cross section of Thomson scattering
$\sigma_{\rm T}$ and the scale factor $a$ normalised to the present
epoch.  The continuity equation for baryons gives the peculiar
velocity of baryons ${{\bm v}_{{\rm b{\bm k}}}}=-i({\bf k}/{k^2}) \dot
\delta_{{\rm b\bf k}}$ where the dot represents the derivative with respect to conformal time. 
Finally, the Doppler anisotropy is thus given by
\begin{equation}
T_D({\bf\hat{n}})
=T_{\rm cmb}\int_0^{\eta_0} d\eta \dot{\tau}e^{-\tau}\int
\frac{d^3k}{(2\pi)^3}
\frac{\dot \delta_{\rm b}}{k^2}
\sum_{\ell} \sqrt{4 \pi (2 \ell +1)} (-i)^\ell  
\frac{\partial}{\partial\eta} j_\ell [k(\eta_0-\eta)]
Y_\ell ^0(\hat{\bm n}).
\label{eq:doppler-map}
\end{equation}
where as above we have taken the Fourier expansion of $\delta_{\rm b}$ with Rayleigh's formula.



During reionisation, CMB polarisation is produced from
the quadrupole component of CMB temperature anisotropy by
Thomson scattering. 
The CMB polarisation can be decomposed into $E$ and $B$-modes with
electric- and magnetic-like parities, respectively.  
We focus on the
dominant modes generated by scalar perturbations.
According to the Boltzmann equations for CMB, 
the scalar perturbations produce only $E$-modes
which are given by \citep{hu-white} 
\begin{equation}
E( \hat {\bm n}) =
\sum _{\ell\; m}  (-i)^\ell \sqrt{ \frac{4 \pi}{2 \ell +1}}
\int {\frac{d^3 k}{(2 \pi)^3}} E^{(0)}_\ell  
Y_\ell ^m (\hat {\bm n}),
\label{eq:emode-map}
\end{equation}
\begin{eqnarray}
{\frac{E^{(0)}_\ell(\eta_0,k)}{2\ell+1}} =   - {\frac{3}{2}}
\sqrt{\frac{(\ell+2)!}{(\ell-2)!}} 
\int_0^{\eta_0} d\eta  \dot\tau e^{-\tau} 
P^{(0)} \frac{j_\ell(k(\eta_0-\eta))}{(k(\eta_0-\eta))^2},
\label{eq:emode-inte}
\end{eqnarray}
where $P^{(0)}$ is the $m=0$ source term due to Thomson scattering. It
is related to the initial gravitational potential $\Phi_0$ via the
transfer function $D_{\rm E}(k, \eta)$, $P^{(0)}=D_{\rm E}(k, \eta )
\Phi_0$; this is detailed in the appendix of \citet{tashiro-polar-2008}.



\subsection{Cross-correlation between 21~cm and CMB}\label{subsec:cross}

The angular power spectrum is defined as the average of the spherical
harmonic coefficients $a _{\ell m}$ over the $(2 \ell +1)$ $m$-values,
$
C_\ell = \sum _m {\langle |a _{\ell m} |^2  \rangle}/{(2 \ell +1)},
$
where the $a _{\ell m}$ are defined for an arbitrary sky map $f( {\bm \hat
  n})$ as  
$
f( {\bm \hat n}) = \sum_{\ell m} a _{\ell m} Y_\ell ^m.
$

From Eqs.~(\ref{eq:21cm-map}) and (\ref{eq:doppler-map}), 
the cross-correlation between the 21-cm line fluctuations 
and the CMB Doppler temperature anisotropy
can be written as
\begin{eqnarray}
C^{\rm 21-D}_\ell (z_{\rm obs}) 
&=&
-\frac{2}{3 \pi}\int_0^\infty k^2dk
\int_0^{\eta_0} d\eta
\left[4 \overline{x}_{\rm H}(z_{\rm obs})
D_{\rm b} (k,\eta_{\rm obs}) k^2 P_{\Phi}( k)
-3 \overline{x}_e(z_{\rm obs}) P_{x \Phi} \right]
\nonumber \\
&& \quad \times
j_\ell [k(\eta_0-\eta_{\rm obs})]j_\ell[k(\eta_0-\eta)]
\frac{\partial}{\partial\eta} \dot{\tau}e^{-\tau} \dot D_{\rm b} (k,\eta),
\label{eq:cross-21T-0}
\end{eqnarray}
where $P_{\Phi}$ and $P_{x \Phi}$ are the power spectra of the initial
gravitational potential and the cross-correlation between the
gravitational potential and the fluctuations of the ionised fraction,
respectively.  The function $D_{\rm b} (k,\eta)$ relates
$\delta_{\rm b}$ to the initial gravitational potential $\Phi_0$ as
$\delta_{\rm b}(k, \eta) = k^2 D_{\rm b} (k, \eta) \Phi_0(k)$, and we have set 
$F\langle \mu^2 \rangle = 1/3$ for the matter dominated epoch. 
We can
simplify Eq.~(\ref{eq:cross-21T-0}) by using the approximation for 
$\ell \gg 1$:
$2 \int_0^\infty dk P(k)j_l(kr)j_l(kr') / \pi \approx 
P\left(k={l/r} \right)\delta(r-r')/l^2$.  
We finally obtain
\begin{eqnarray}
\ell^2 C^{\rm 21-D}_\ell (z_{\rm obs}) =
-\frac{1}{3} \left(\frac{\ell}{r_{\rm obs}} \right)^2
\left[4 \overline{x}_{\rm H}(z_{\rm obs})
D_{\rm b} (k,\eta_{\rm obs}) \left(\frac{\ell}{r_{\rm obs}} \right)^2 
P_{\Phi} \left(\frac{\ell}{r_{\rm obs}} \right)
-3 \overline{x}_e(z_{\rm obs}) P_{x \Phi} \left(\frac{\ell}{r_{\rm obs}}  ,z_{\rm obs}\right)
\right]
\frac{\partial}{\partial\eta'} \dot{\tau}e^{-\tau} \dot D_{\rm b} (k,\eta) 
|_{\eta=\eta_{\rm obs}}.
\label{eq:cross-21T}
\end{eqnarray}
Eq.~(\ref{eq:cross-21T}) involves two terms. One involves $P_{\Phi}$ 
and is the homogeneous ionisation term. The other term involves $P_{x\Phi}$ 
and is the bias term. 
The homogeneous term corresponds to the anti-correlation part of the signal. In
over-dense regions, the 21-cm emission is strong due to the
large amounts of hydrogen ($\delta_{21} >0$); while the CMB
temperature is lower due to the Doppler shift ($\delta_{\rm Doppler} <
0$).  The bias term in turn shows the positive correlation part of the
signal.  In over-dense regions, ionising sources are numerous and the
quantity of neutral hydrogen is small. 
Therefore, the 21-cm emission in over-dense regions is weaker than
the background emission ($\delta_{21} < 0$).

The cross-correlation between 21~cm line fluctuations and $E$-modes
was studied in detail by \citet{tashiro-polar-2008}. We provide here
the basic equation 
\begin{equation}
C_\ell ^{E-21} 
=
-\frac{3}{\pi} T_0 \sqrt{\frac{(\ell+2)!}{(\ell-2)!}} 
\int {dk } \int d \eta k^2 \dot \tau e^{-\tau} 
D_E  (k, \eta )\left[ \frac{4}{3} 
(1- \bar x_e) P_{\Phi \delta_{\rm b}}- \bar x_e P_{x \Phi}
\right] \frac{j_\ell (k( \eta_0 -\eta_{\rm obs})) j_\ell  (k( \eta_0
-\eta))}{(k( \eta_0 -\eta))^2}.
\label{eq:cross-21E}
\end{equation}
where $P_{\Phi \delta_{\rm b}}$ is the power spectrum of 
the cross-correlation between the gravitational potential
and the baryon density fluctuations.
According to the cosmological linear perturbation theory \citep[e.g.][]{kodama-sasaki-1984}
the power spectrum $P_{\Phi \delta_{\rm b}}$ can be written in terms
of the initial power spectrum of the gravitational potential $P_{\Phi}$ as
$
P_{\Phi \delta_{\rm b}} = k^2 D_{\rm b}(k, \eta ) P_{\Phi}.
$
The function $D_E (k, \eta )$ exhibits an oscillatory
behavior and it can be decomposed as well into a
homogeneous-ionisation and a bias terms. 
However, their signs depend on
$D_E$.

\subsection{Reionisation model}

Cross-correlations between 21~cm and CMB in Eqs.~(\ref{eq:cross-21T})
and (\ref{eq:cross-21E}) involve two power spectra $P_{\Phi}$ and
$P_{x \Phi}$.  While $P_{\Phi}$ is computed using the WMAP
cosmological parameters, $P_{x \Phi}$ depends on the reionisation
process.  Although the latter is not well-known, we can reasonably
expect that ionising sources are formed in dense regions and that they
ionise the surrounding medium with an efficiency that depends on the
density of the medium.  Therefore, we can distinguish two possible
cases: One where ionised fluctuations and matter over-densities
coincide, and the other where ionised fluctuations and matter density
are anti-biased \citep[e.g.][]{benson-2001}. Following
\citet{alvarez-komatsu}, we assume that the fluctuations of the
ionised fraction are associated with the matter density contrast using
the Press-Schechter description \citep[][]{ps-1974}.  As a result, the
power spectrum $P_{x \Phi}$ is given by
\begin{equation}
\bar x_e P_{x \Phi} =
- \bar x_{\rm H} \ln \bar x_{\rm H} 
[\bar b_{\rm h}-1-f] D_{\rm m}(k, \eta ) k^2 P_{\Phi},
\end{equation}
where $D_{\rm m}$ is the transfer function of matter (both dark and
baryonic), $\bar b_{\rm h}$ is the average bias of dark matter halos
more massive than the minimum mass of the source of
ionising photons $M_{\rm min}$
\begin{equation}
\bar b_{\rm h} = 1 + \sqrt{\frac{2}{\pi}} \frac{e^{-\delta_c ^2 /2
\sigma^2(M_{\rm min})}}{f_{\rm coll} \sigma (M_{\rm min})},
\end{equation}
where $\sigma(M)$ is the variance of the density fluctuations smoothed
with a top-hat filter of the scale corresponding to a mass $M$, and
$f_{\rm coll}$ is the fraction of matter collapsed into halos with
$M>M_{\rm min}$.  In this paper, we choose $M_{\rm min}$ such that the
halo virial temperature is $T_{\rm vir}(M_{\rm min}) = 10^4$ K. This
choice corresponds to the assumption that the ionising sources form in dark
matter halos where the gas cools efficiently via atomic cooling. 
The parameter $f$ describes the reionisation regime we
are interested in. For $f=0$, we are in the ``photon-counting limit''
case where recombinations are not important and where the progress of
the reionisation depends on the number of ionising photons only.  The
over-dense regions contain more collapsed objects which are sources of
ionising photons.  Therefore, in this case, ionisation in 
over-dense regions is easier than in  under-dense regions.  On the
contrary, $f=1$ indicates the ``Str\"{o}mgren limit" case where 
ionisation is balanced by recombination.  Although the
over-dense regions contain more sources of ionising photons, the
recombination rate in  over-dense regions is higher than in
under-dense regions. Hence, over-dense regions in the $f=1$ case
have a lower ionised fraction than in the $f=0$ case
\citep[for details, see][]{alvarez-komatsu}.

Finally, in order to calculate the cross-correlation, we need the
evolution of the mean ionised fraction for which we use a
simple parameterisation based on two key quantities, the reionisation
redshift (defined as the redshift at which the ionised fraction equals
0.5), $z_{\rm re}$, and the reionisation duration, $\Delta z$,
\begin{equation}
\bar x_e (z) = \frac{1}{1 + \exp[(z-z_{\rm re})/\Delta z]}.
\label{eq:xh}
\end{equation}

\section{Cross-correlation power spectrum}\label{sec:crosspower}

\begin{figure}
  \begin{tabular}{cc}
   \begin{minipage}{0.5\textwidth}
  \begin{center}
    \includegraphics[keepaspectratio=true,height=50mm]{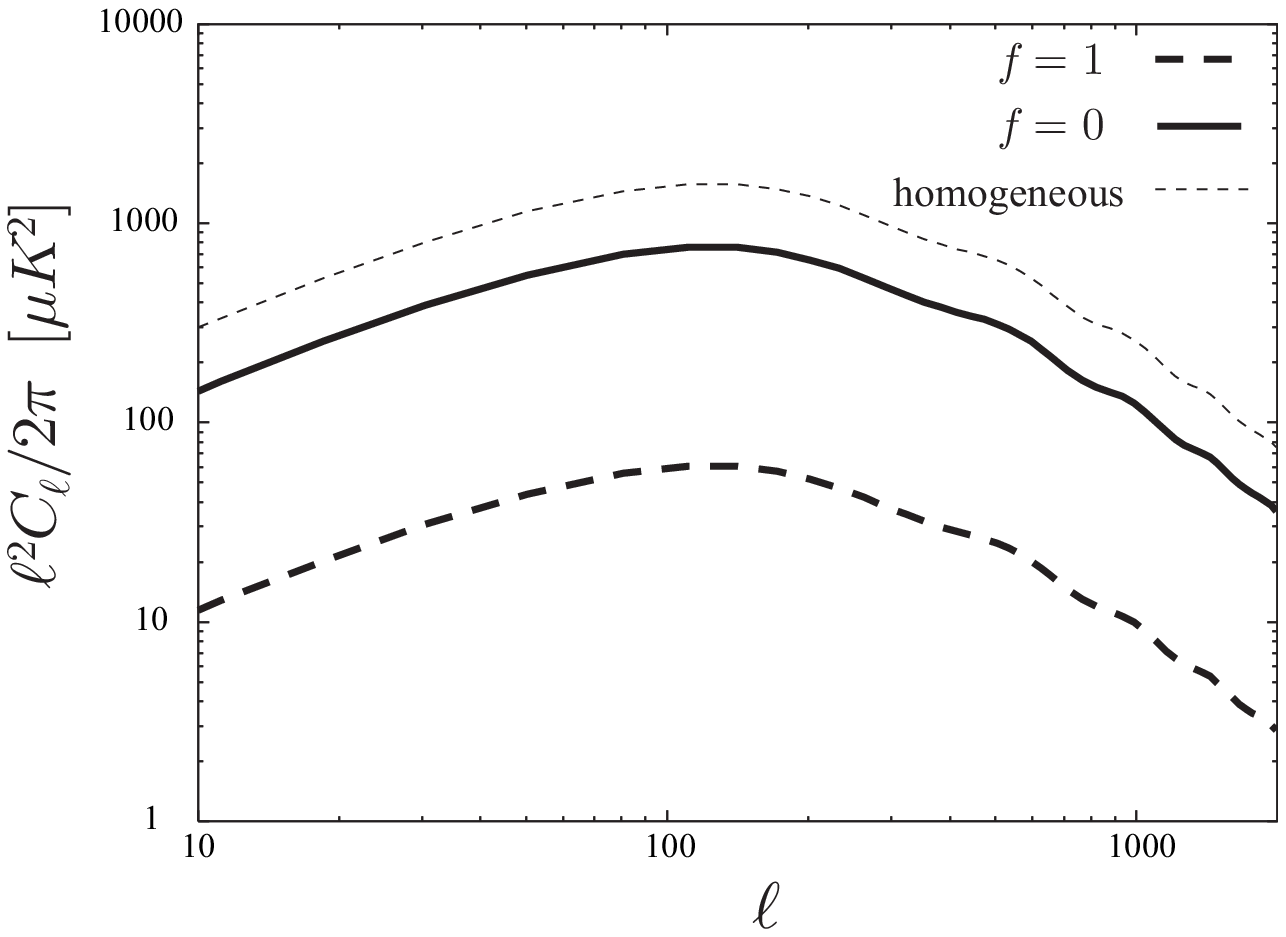}
  \end{center}
  \end{minipage}
   \begin{minipage}{0.5\textwidth}
  \begin{center}
    \includegraphics[keepaspectratio=true,height=50mm]{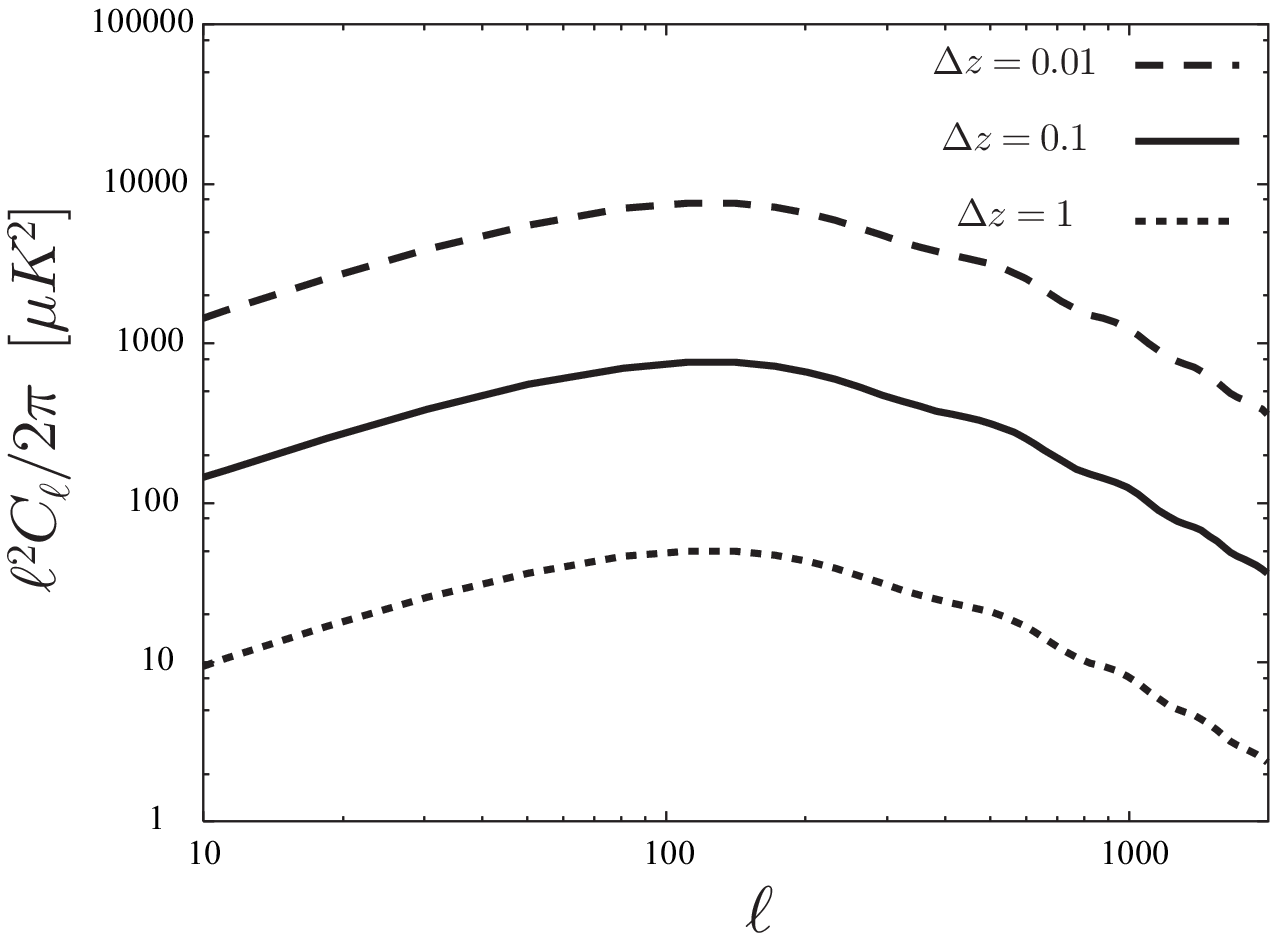}
  \end{center}
   \end{minipage}
  \end{tabular}
  \caption{The cross-correlation between 21-cm fluctuations and CMB
    Doppler temperature anisotropy.  In the left panel, we show the
    dependence on the reionisation model.  We set $z_{\rm re} = 10$, $
    \Delta z = 0.1$ and $z_{\rm obs} = 10$.  The solid line represents
    the $f=0$ case where we do not take into account the recombinations. 
    The dashed line is for the $f=1$ case where
    recombinations and ionisations are balanced.  The thin dotted line
    represents the homogeneous term where we do not consider the
    fluctuations of the ionised fraction $\delta_x$.  In the right
    panel, the dependence on the reionisation duration is shown for
    $z_{\rm re} = 10$, $ f=0$ and $z_{\rm obs} = 10$.  The dashed,
    solid and dotted lines are for $ \Delta z = 0.01$, $ \Delta z =
    0.1$ and $ \Delta z = 1$, respectively.  }
  \label{fig:T21re10}
\end{figure}

In the left panel of Fig.~\ref{fig:T21re10}, we show the power
spectrum of the cross-correlation between the 21-cm line fluctuations and
the Doppler anisotropy. For this computation, we set the reionisation
redshift and duration as $z_{\rm re}=10$, $\Delta z=0.1$ and we take
$z_{\rm obs}=10$. We explore both the photon-counting-limit case ($f=0$)
and the Str\"{o}mgren-limit case ($f=1$). In both cases the
cross-correlation has a positive sign. As mentioned earlier, more
fluctuations are produced in the photon-counting-limit case than in the
Str\"{o}mgren-limit case. The amplitude of the power spectrum with
$f=0$ is thus larger than that with $f=1$. 

The cross-correlation signal has two different contributions
with opposite signs as shown in Sec.~\ref{subsec:cross}. 
One is associated with the bias term and the other is with the homogeneous
term. For reference, we plot the homogeneous ionisation part as the
thin line in the left panel of Fig.~\ref{fig:T21re10}.  
At high redshifts ($z>15$), since the average bias is high,
the bias part dominates the homogeneous part
as shown in the model of \citet{alvarez-komatsu} where
they have taken $z_{\rm re}=15$ and $z_{\rm obs}=15$.
However, at low redshifts ($z<15$), since the bias is the order of 1,
the bias term is comparable to the homogeneous part.
Therefore, in our reionisation model where $z_{\rm re}=10$ and $z_{\rm obs}=10$, 
cancellation occurs in the total signal.
Subsequently, the total amplitude of the cross-correlation ends up 
smaller than that in the homogeneous ionisation part.

The right panel of Fig.~\ref{fig:T21re10} exhibits 
the dependence of the cross-correlation power spectrum 
on the reionisation duration for the
case with $z_{\rm re}=10$, $z_{\rm obs}=10$ and $f=0$.  When the
reionisation time is fixed, the shorter the duration the larger the
amplitude of the power spectrum. As a matter of fact, long duration of
the reionisation increases the integration range over $\eta$ in
Eq.~(\ref{eq:cross-21T-0}) and thus causes cancellation of the
correlation due to phase gap between the density
and velocity fluctuations.  Note that, according
to \citet{alvarez-komatsu}, the instantaneous reionisation gives an infinite
signal (Eq.~\ref{eq:cross-21T}).  However, Eq.~(\ref{eq:cross-21T}) is
obtained using the Limber approximation which is no more valid in a
short duration reionisation.  We therefore perform an exact
calculation of the cross-correlations from Eq.~(\ref{eq:cross-21T-0}).

The cross-correlation between 21-cm line fluctuations and CMB $E$-mode
polarisation was studied in detail in \citet{tashiro-polar-2008}.  The
angular power spectrum depends on the polarisation source term
$P(k_{\rm obs})$, namely the quadrupole term of the CMB, at $z_{\rm
  obs}$ where $k_{\rm obs}$ satisfies $k_{\rm obs} = \ell / (\eta_0 -
\eta_{\rm obs})$.  Accordingly, the angular power spectrum exhibits
its first peak at a multipole $\ell<10$ which corresponds to the
angular separation of the quadrupole at $z_{\rm obs}$. The free
streaming of the quadrupole at redshifts higher than $z_{\rm obs}$
produces oscillations at higher $\ell$ modes ($\ell >10$). These oscillations
are increasingly damped by larger reionisation durations $\Delta z$.
In addition and similarly to the cross-correlation between 21-cm line
fluctuations and CMB Doppler temperature anisotropy, the parameter $f$
affects the amplitude of the cross-correlation with the $E$-modes.
The $f=0$ case produces more fluctuations than the $f=1$ case, and
thus a larger overall amplitude.

\section{Detection of the cross-correlation signals}

For computation of the SN ratio, evaluating the noise power 
is crucial.
Especially, the estimation of the experimental
noise power spectrum for each observation strategy is an important factor
of the noise power spectrum.
Here, we introduce the parameterisation of the experimental
noise for the various planned observation:
LOFAR, MWA, SKA.
Then, we calculate the SN ratio for the 21~cm cross-correlation with CMB Doppler
temperature and CMB $E$-mode polarisation which are given by Eqs.~(\ref{eq:cross-21T}) 
and (\ref{eq:cross-21E}), respectively.

\subsection{Noise power spectrum}\label{sec:noise}

In order to evaluate the noise power spectrum, we neglect the foregrounds. 
Under this assumption, the noise power spectrum of the signal from reionisation 
consists of the experimental noise power spectrum and of the power spectrum of
primary CMB.

For the CMB observation, we consider the {\it Planck} configuration. 
In this case, compared with the CMB signal,
the experimental noise is very small on scales of interest.
Therefore, we neglect the experimental noise power spectrum.
This assumption gives the noise for the CMB Doppler
temperature anisotropy as $N_\ell ^{D} = C_\ell ^{T}$  and 
for $E$-mode polarisation from reionisation
as $N_\ell ^{E} = C_\ell ^{E}$ 
where $C_\ell ^{E}$ is the primary CMB $E$-modes. 


For the 21-cm fluctuations,
the dominant signal of the 21-cm line on large scales
is that of reionisation.
Therefore, we can assume that the noise spectra of the 21-cm
fluctuations consist of the experimental noise power spectra only.
According to \citet{zaldarriaga-furlanetto-hernquist}, the power
spectrum of the experimental noise of the 21-cm observations at a
wavelength $\lambda$ cm is given by
\begin{equation}
{\ell^2 N_\ell ^{21} \over 2 \pi}= 
\left ( {\ell  \over 100} \right)^2  N_{100}, 
\end{equation}
where $N_{100}$  is a normalised
noise power spectrum which is written as
\begin{equation}
N_{100}= {1 \over t_{\rm obs} \Delta \nu} \left( {100 \ell_{\rm
max} \over 2 \pi} {\lambda^2 \over A/T} \right)^2.
\end{equation}
Here $\Delta \nu$ is the bandwidth, $t_{\rm obs}$ is the total
integration time, $A/T$ is the sensitivity (an effective area divided
by the system temperature) and $\ell_{\rm max}= 2 \pi {D \over
  \lambda}$ is the maximum multipole associated with the length of the
baseline $D$. In Table \ref{table:design}, we summarise the main
characteristics of the present designs of MWA
\citep{bowman-MWA-2006,lidz-zahn-2008}, LOFAR \citep{lofar-jelic-2008}
and SKA \citep{alvarez-komatsu} and calculate $\sqrt{N_{100}}$ for the
observation wavelength corresponding to an observing redshift $z_{\rm
  obs}=10$ matching the present reionisation limits.  In the table,
LOFAR-1 and LOFAR-3 stand for two cases one with a single
  observed field, LOFAR-1, and the second with the three observed
  fields, LOFAR-3.  For reference, we consider an ideal experiment
which we refer to as ``super SKA'' with a sensitivity 10 times that of
SKA and a field of view twice as large as SKA's.

\begin{table}
 \begin{center}
  \begin{tabular}{|c|c|c|c|c|c|c|}
    \hline
   & $f_{\rm sky}$ & $\Delta \nu ~ ({\rm MHz})$ 
   &  $t_{\rm obs}$  &  $A/T$ ~(${\rm m}^2/{\rm K}$) & $ D ~({\rm
      Km})$  &$ \sqrt{N_{100}}~ ({\rm \mu K})$ \\
    \hline
    MWA   & 0.02  & 6 & 1000~{\rm hour}  &  13  & 1.5 & $5600$ \\
    \hline
    LOFAR-1& 0.0024 & 1 & 800~{\rm hour}   & 108   & 2 & $1200$  \\
    \hline
    LOFAR-3& 0.007 & 1 & 1500~{\rm hour}   & 108   & 2 & $900$  \\
    \hline
    SKA   & 0.009  & 1 & 1 ~{\rm month}   &  1000  & 1 &$140 $   \\
    \hline
    super SKA   & 0.018  & 1 & 1 ~{\rm month}   &  1000  & 1 &$70 $   \\
    \hline
  \end{tabular}
 \end{center}
 \caption{The current designs of 21~cm experiments. The estimated
   $\sqrt{N_{100}}$ is computed for the observation wavelength 
   which corresponds to $z_{\rm obs}=10$.}
\label{table:design}
\end{table}

\subsection{Results}

\begin{figure}
  \begin{tabular}{cc}
   \begin{minipage}{0.333\textwidth}
  \begin{center}
    \includegraphics[keepaspectratio=true,height=47mm]{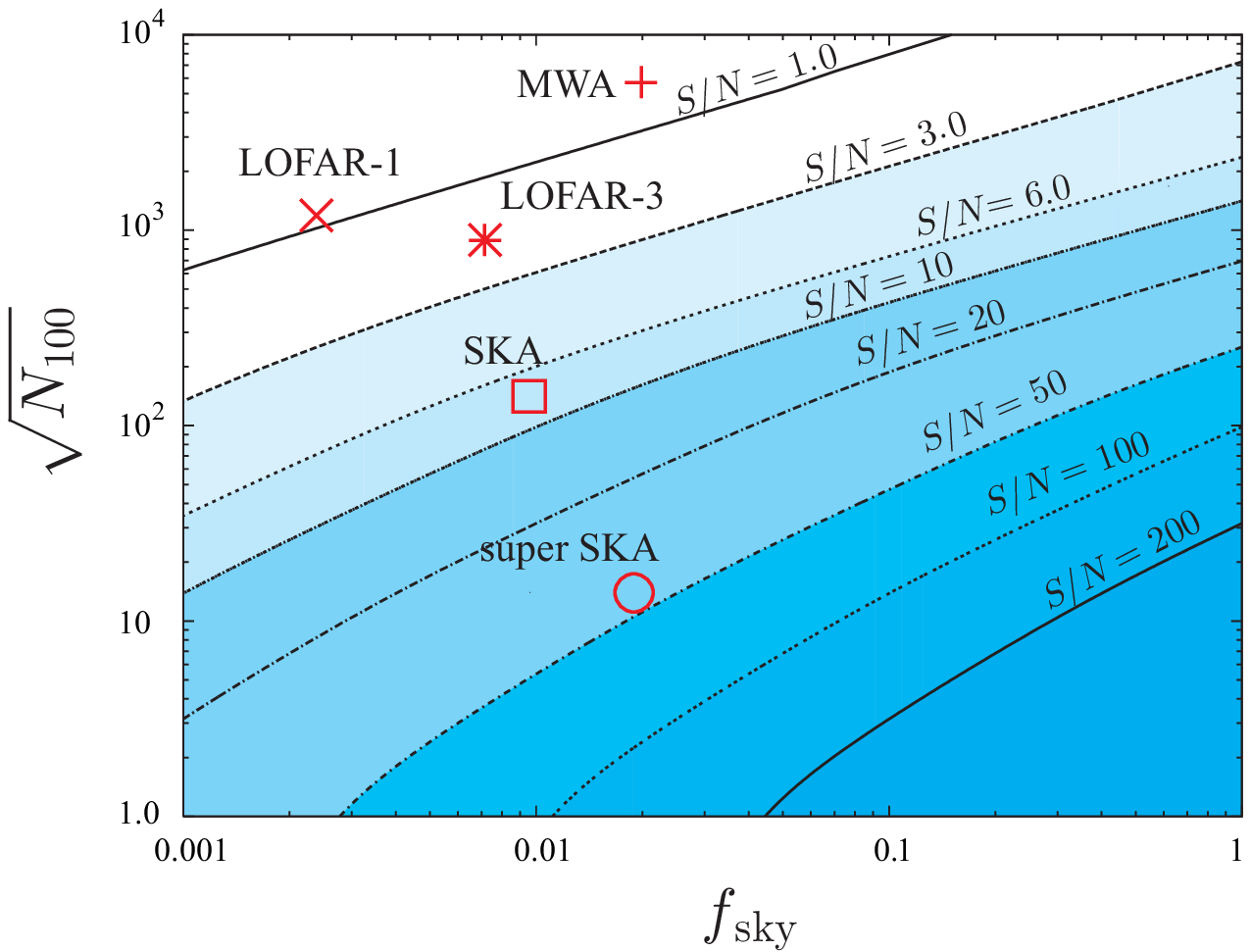}
  \end{center}
  \end{minipage}
   \begin{minipage}{0.333\textwidth}
  \begin{center}
    \includegraphics[keepaspectratio=true,height=47mm]{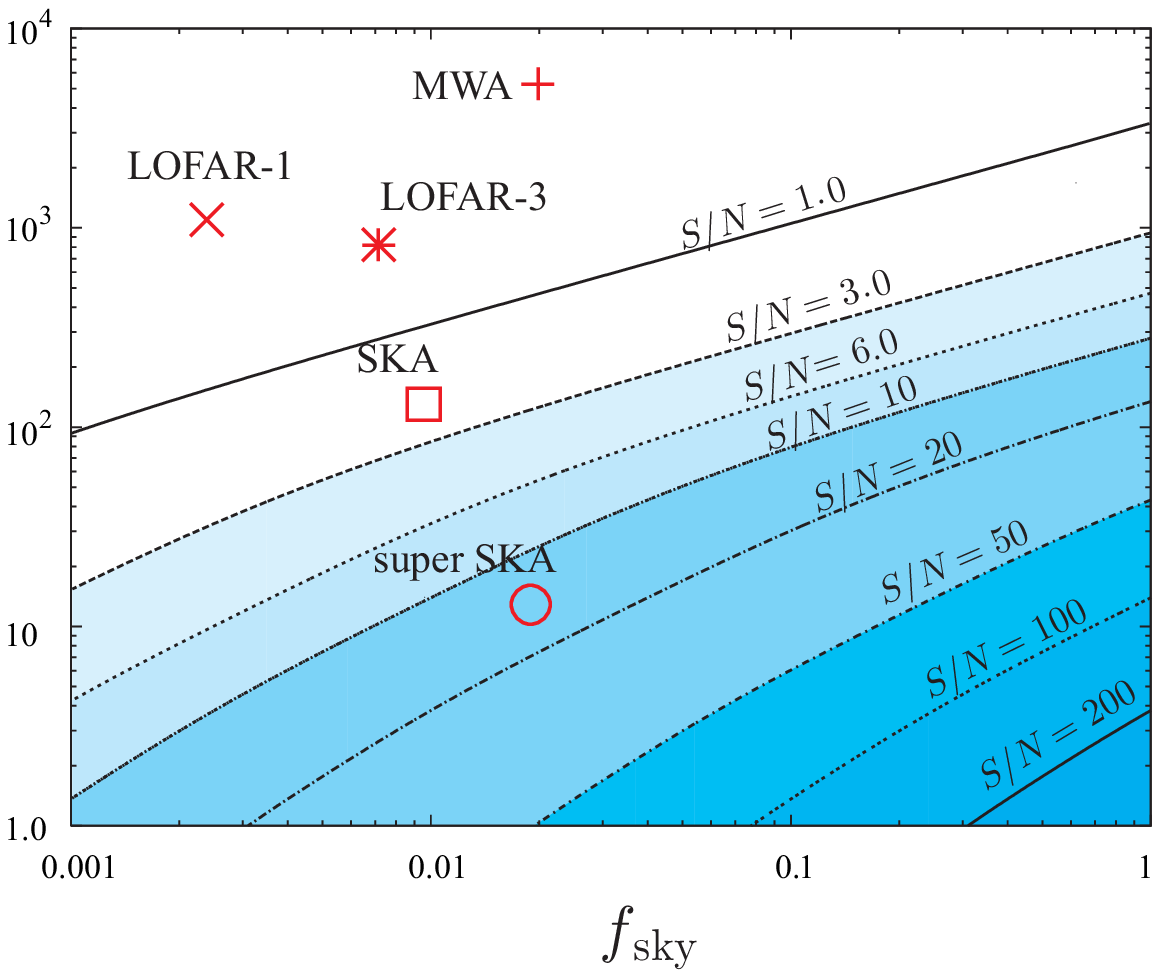}
  \end{center}
   \end{minipage}
   \begin{minipage}{0.333\textwidth}
  \begin{center}
    \includegraphics[keepaspectratio=true,height=47mm]{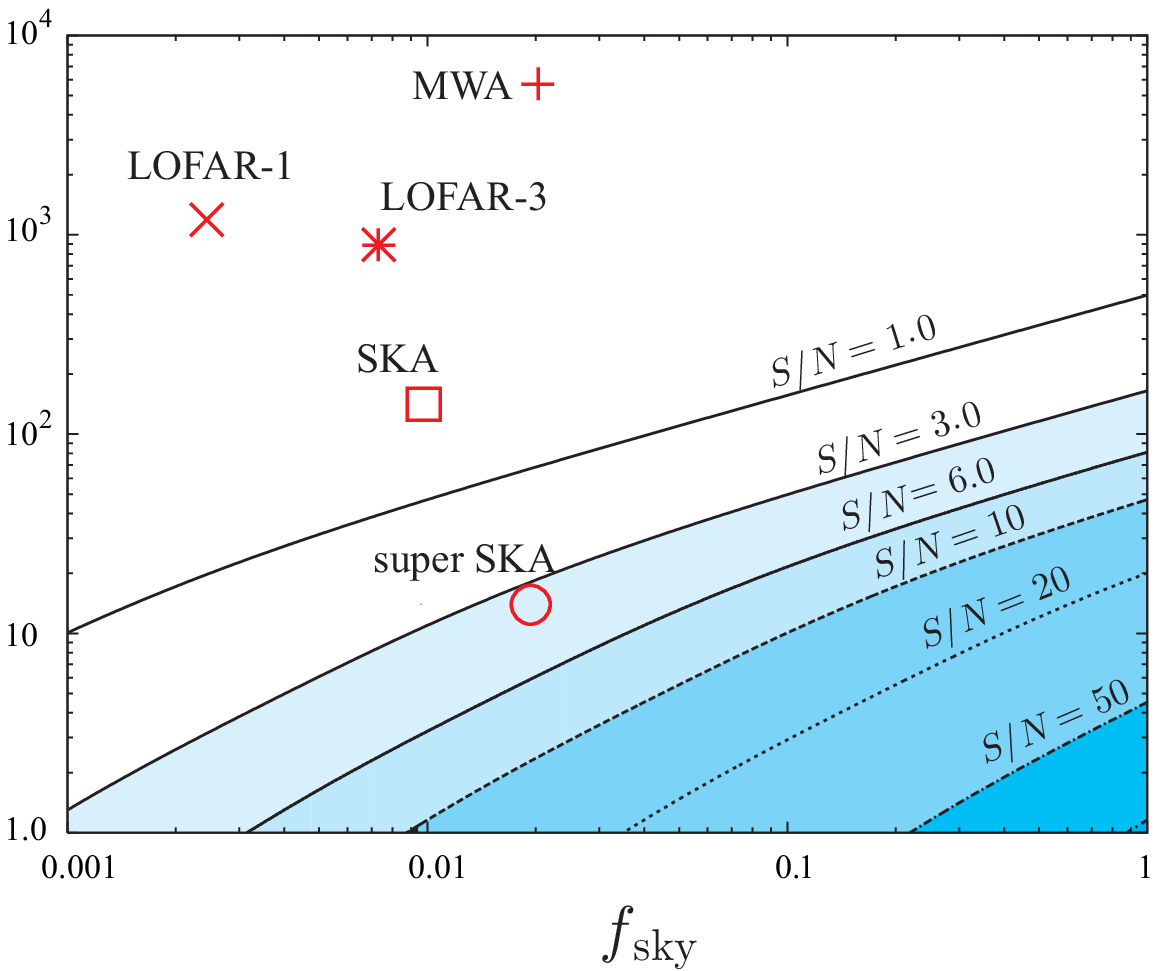}
  \end{center}
   \end{minipage}
  \end{tabular}
  \caption{SN ratio of the 21-cm cross-correlation with the CMB
    Doppler anisotropy for different reionisation durations.  In all
    panels, the SN ratio is given as a function of the sky fraction
    $f_{\rm sky}$ and the normalised noise power spectrum $N_{100}$.
    In all panels, we set $z_{\rm obs}=10$ and $z_{\rm re}=10$.  From
    left to right, the reionisation durations are set to $\Delta z=
    0.01$, $\Delta z= 0.1$ and $\Delta z= 0.5$.}
  \label{fig:21Tcross}
\end{figure}

\begin{figure}
  \begin{tabular}{cc}
   \begin{minipage}{0.333\textwidth}
  \begin{center}
    \includegraphics[keepaspectratio=true,height=47mm]{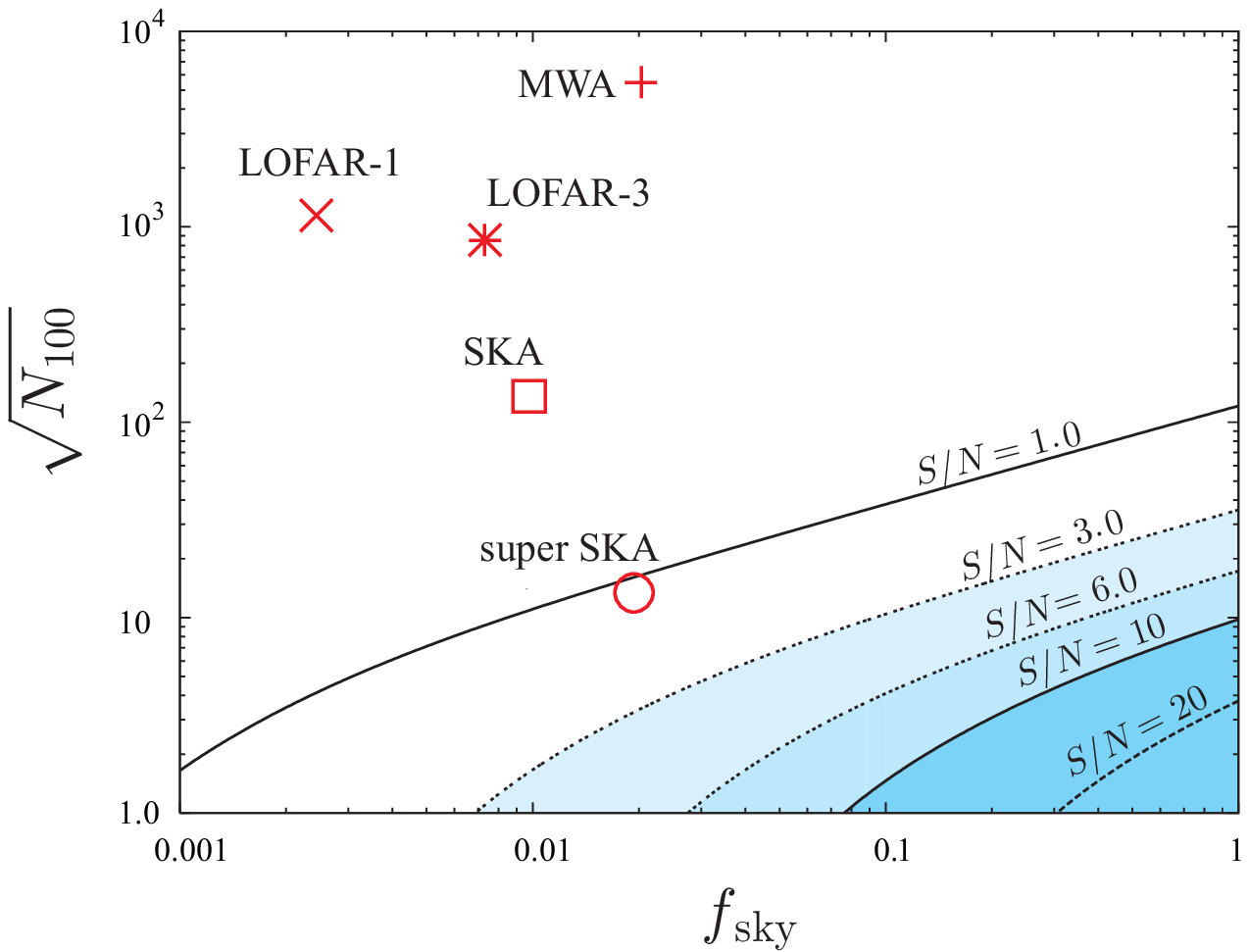}
  \end{center}
  \end{minipage}
   \begin{minipage}{0.333\textwidth}
  \begin{center}
    \includegraphics[keepaspectratio=true,height=47mm]{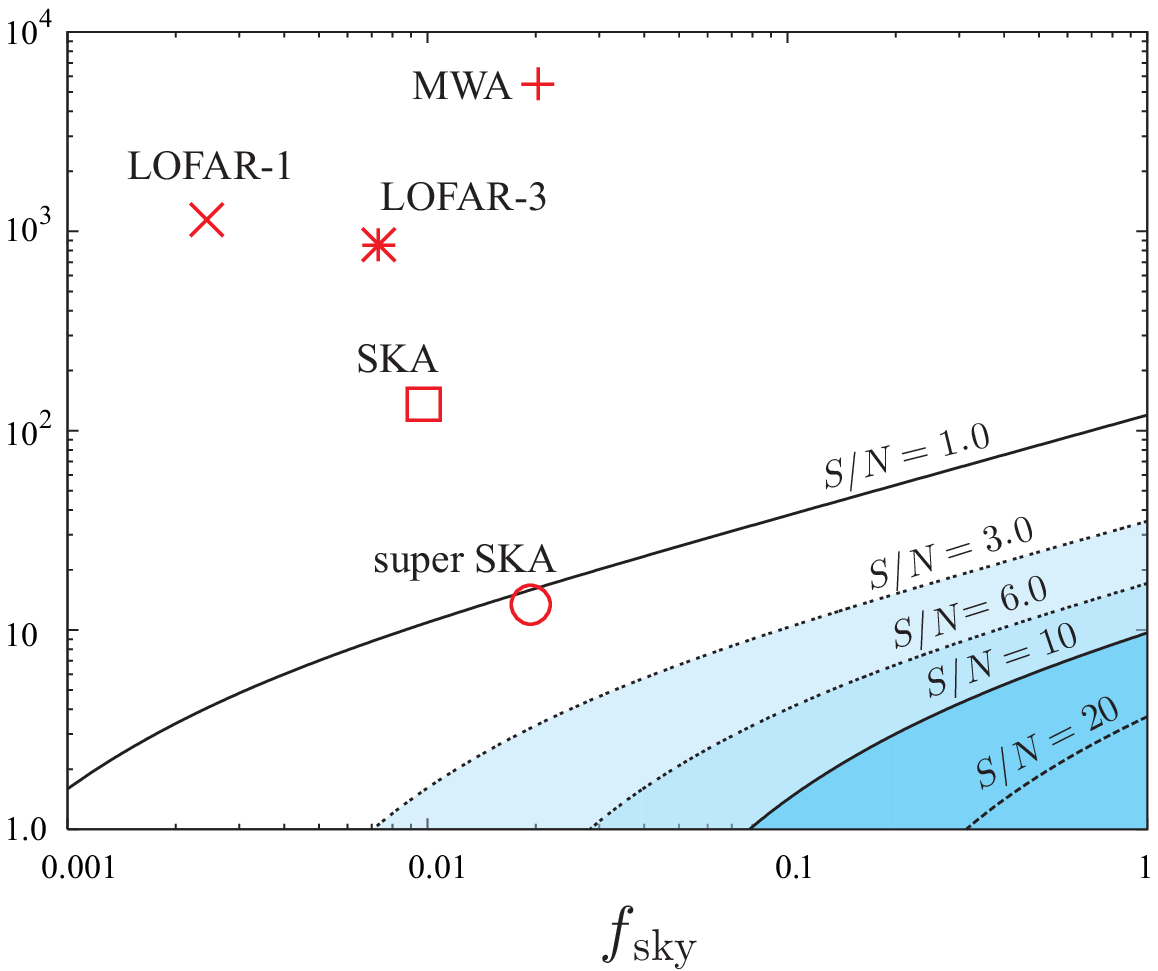}
  \end{center}
   \end{minipage}
   \begin{minipage}{0.333\textwidth}
  \begin{center}
    \includegraphics[keepaspectratio=true,height=47mm]{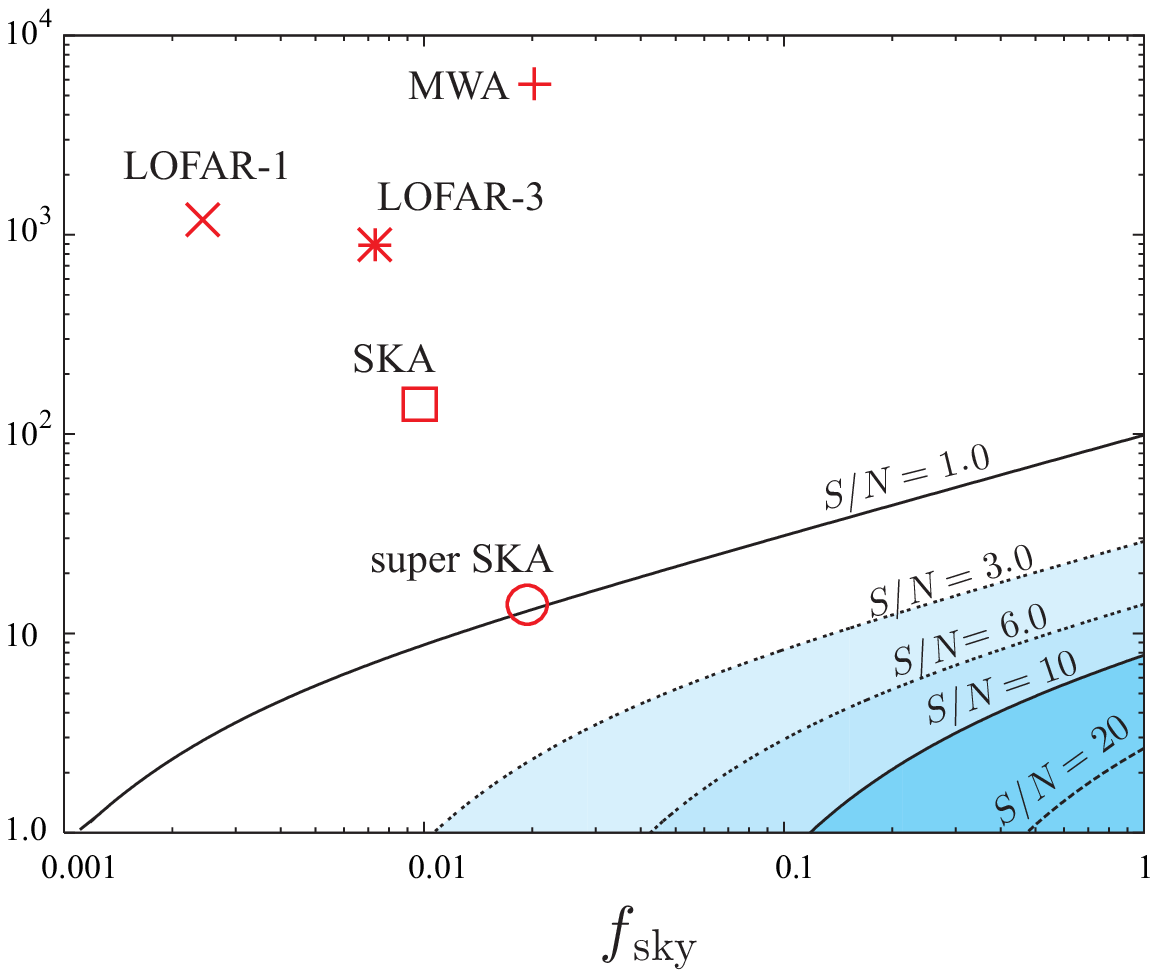}
  \end{center}
   \end{minipage}
  \end{tabular}
  \caption{SN ratio of the 21-cm cross-correlation with the CMB
    $E$-modes for different reionisation durations. In all panels, the
    SN ratio is given as a function of the sky fraction $f_{\rm sky}$
    and the normalised noise power spectrum $N_{100}$.  In all panels,
    we set $z_{\rm obs}=10$ and $z_{\rm re}=10$.  From left to right,
    the reionisation durations are set to $\Delta z= 0.01$, $\Delta z=
    0.1$ and $\Delta z= 0.5$.}
  \label{fig:21Ecross}
\end{figure}

We calculate the SN ratio for the cross-correlation between 21~cm
fluctuations and the CMB Doppler temperature anisotropy
(Fig.~\ref{fig:21Tcross}) and CMB $E$-modes (Fig.~\ref{fig:21Ecross})
for a reionisation model with $z_{\rm re}=10$ and different
reionisation durations.  In both figures, we show the dependence of SN
ratio on $f_{\rm sky}$ and $N_{100}$. From left to right $\Delta z$ is
set to 0.01, 0.1 and 0.5.  In these two-parameter-space figures, we
show the positions of the current experimental designs for 21~cm
observations (see also Table \ref{table:design}).  
Fig. \ref{fig:21Tcross} shows that the cross-correlation between
Planck and LOFAR, in its present configuration, is only sensitive to
an ``instantaneous'' reionisation (with $\Delta z =0.01$). If the
quantity $N_{100}$, expressing the instrumental noise of LOFAR, were
reduced by a factor ten (by improving the sensitivity $T/A$ or
increasing the observation time $t_{\rm obs}$), LOFAR would detect the
cross-correlation signal from the instantaneous reionisation with $S/N
> 3$ for single observation field and $S/N
> 5$ for multi observation fields. 
As shown in Sec.~\ref{sec:crosspower}, the longer the
duration of reionisation $\Delta z$, the smaller the amplitude of the
cross-correlation. As a result, LOFAR becomes insensitive to the
reionisation signal for $\Delta z=0.1$ whereas SKA sees the
signal-to-noise decreasing from $S/N=8$ when $\Delta z=0.01$ to $S/N=2.5$ 
when $\Delta z=0.1$. When the reionisation is longer, $\Delta z = 0.5$, the
cross-correlation signal is detected only by an ideal experiment like
``super SKA''.

\begin{figure}
  \begin{center}
    \includegraphics[keepaspectratio=true,height=80mm]{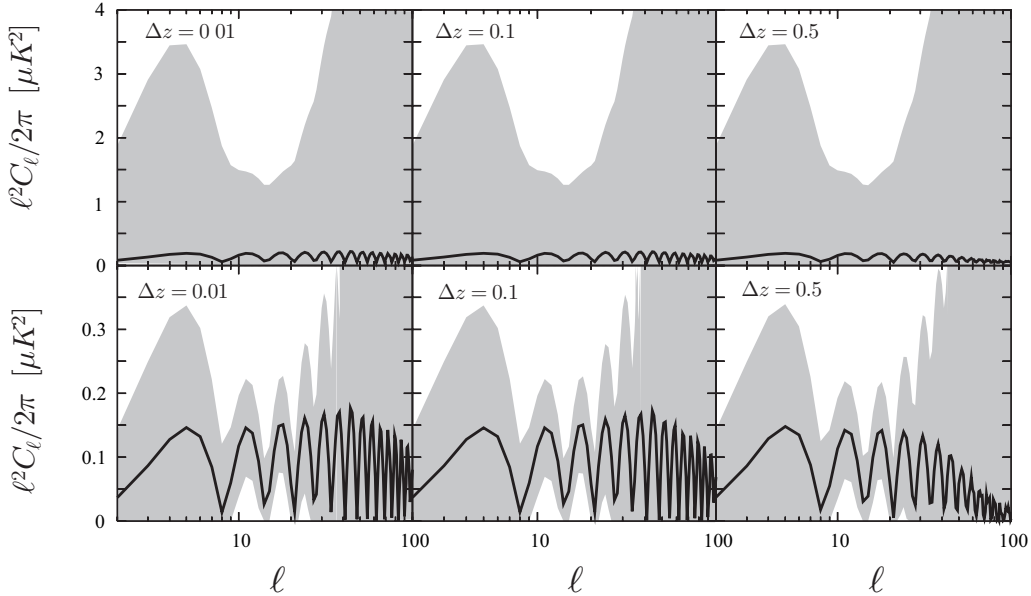}
  \end{center}
  \caption{The 21-cm and $E$-mode cross-correlation signal with the
    estimated errors.  We set $z_{\rm obs}=10$, $z_{\rm re}=10$ and
    $f=0$.  From the left to the right, we take $\Delta z= 0.01$,
    $0.1$ and $0.5$.  The top panels show the cross-correlation error
    for SKA and, the bottom panels are for {\it super SKA}.  The
    cross-correlation signal is the solid line and the error regions
    are represented as the gray zone in each panel.  }
    \label{fig:E21skax100}
\end{figure}

Fig. \ref{fig:21Ecross} shows that the
cross-correlation signal is detected only by an ideal experiment like
``super SKA'' with at most $S/N=1.0$.
We show, in Fig.~\ref{fig:E21skax100}, the cross-correlation power
spectrum between 21~cm and CMB $E$-modes with the errors estimated
from Eq.~(\ref{eq:error}). As mentioned previously, increasing the
duration of reionisation 
damps the power at high
$\ell$s. At those scales, the noise due to CMB signal dominates the
cross-correlation signal making it very difficult to probe the
duration of reionisation (see Fig.~\ref{fig:E21skax100}). 
As a result, 
the SN ratio does not depend on $\Delta z$ as shown in Fig. \ref{fig:21Ecross}.

\begin{figure}
  \begin{center}
    \includegraphics[keepaspectratio=true,height=100mm]{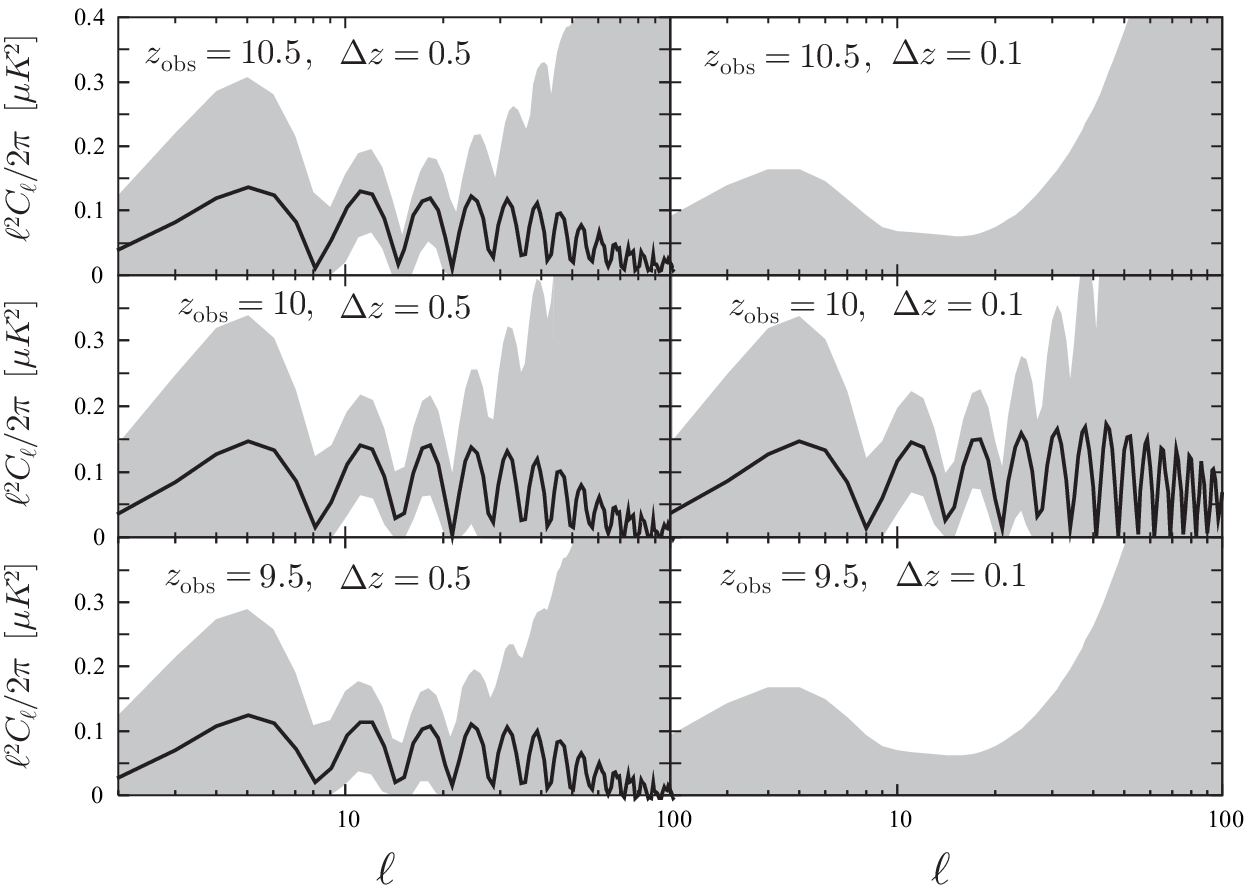}
  \end{center}
  \caption{ The 21-cm and $E$-mode cross-correlation signal with the
    estimated errors, for SKA, at different observing redshifts.  We
    set $z_{\rm re}=10$ and $f=0$ with $\Delta z=0.5$ and $\Delta
    z=0.1$ in the left and right panels, respectively. From top to
    bottom, we set $z_{\rm obs}= 10.5$, $10$ and $9.5$. The
    cross-correlation signal is shown as the solid line and the errors
    are represented as the gray zone in each panel.  }
    \label{fig:E21Z05}
\end{figure}

The amplitude of the cross-correlation gradually increases as the redshift
$z_{\rm obs}$ goes down. The signal reaches its maximum value at
$z_{\rm obs}=z_{\rm re}$ where the ionised fraction is about
one half. Tracing this evolution in the cross-correlation signal with
future radio-interferometer observations may possibly constrain
the duration of reionisation. This is illustrated in
Fig.~\ref{fig:E21Z05}, where we show the cross-correction with the
estimated error at different redshifts in the ideal case of {\it super
  SKA} for two different reionisation durations, $\Delta z =0.1$ and
$\Delta z =0.5$. We show that the signal from the instantaneous
reionisation $\Delta z =0.01$ vanishes before or after
the redshift $z_{\rm re}$,
whereas the signal from a longer duration, $\Delta z =0.5$, does not
disappear.


\begin{figure}
  \begin{tabular}{cc}
   \begin{minipage}{0.5\textwidth}
  \begin{center}
    \includegraphics[keepaspectratio=true,height=50mm]{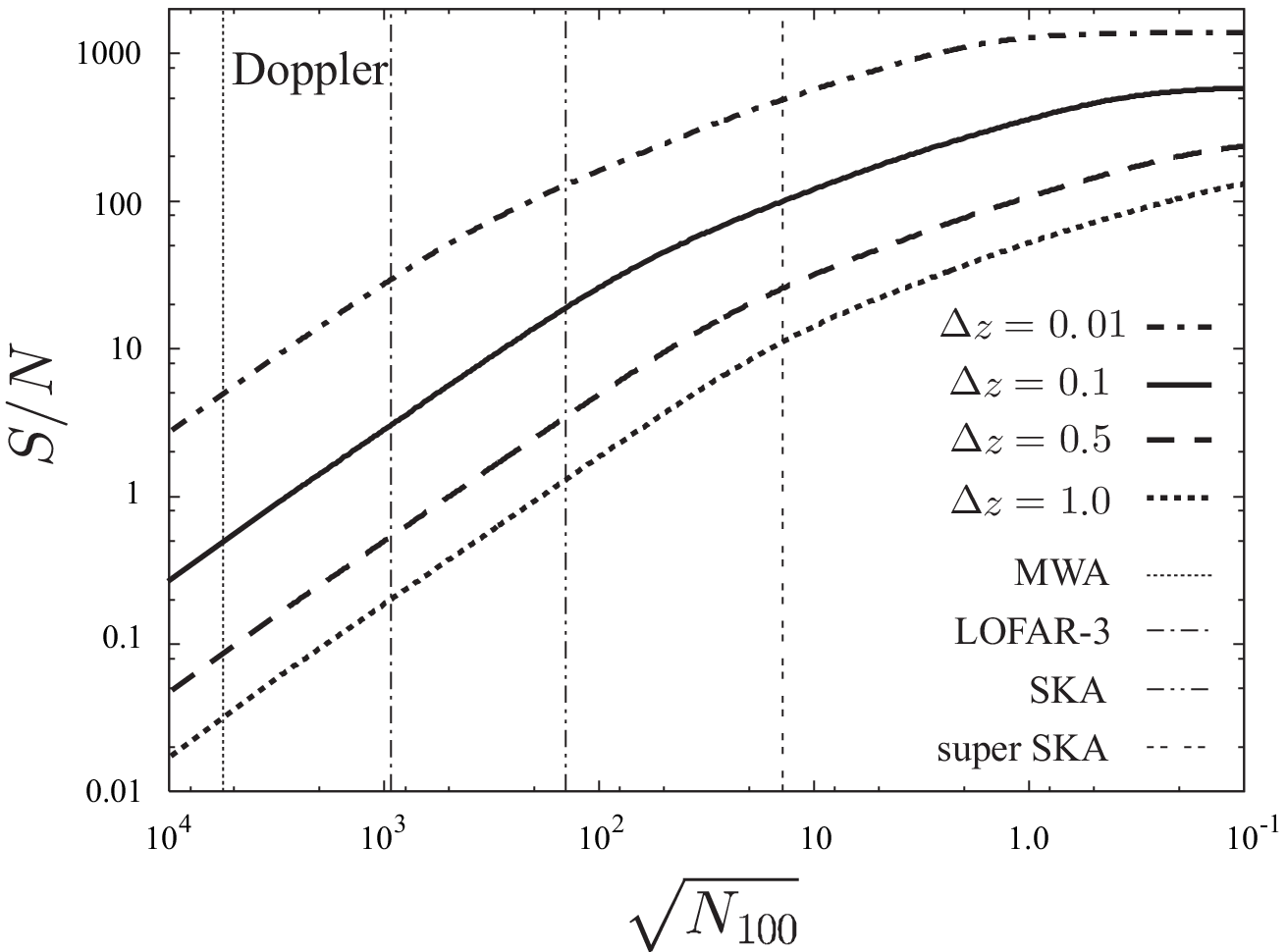}
  \end{center}
  \end{minipage}
   \begin{minipage}{0.5\textwidth}
  \begin{center}
    \includegraphics[keepaspectratio=true,height=50mm]{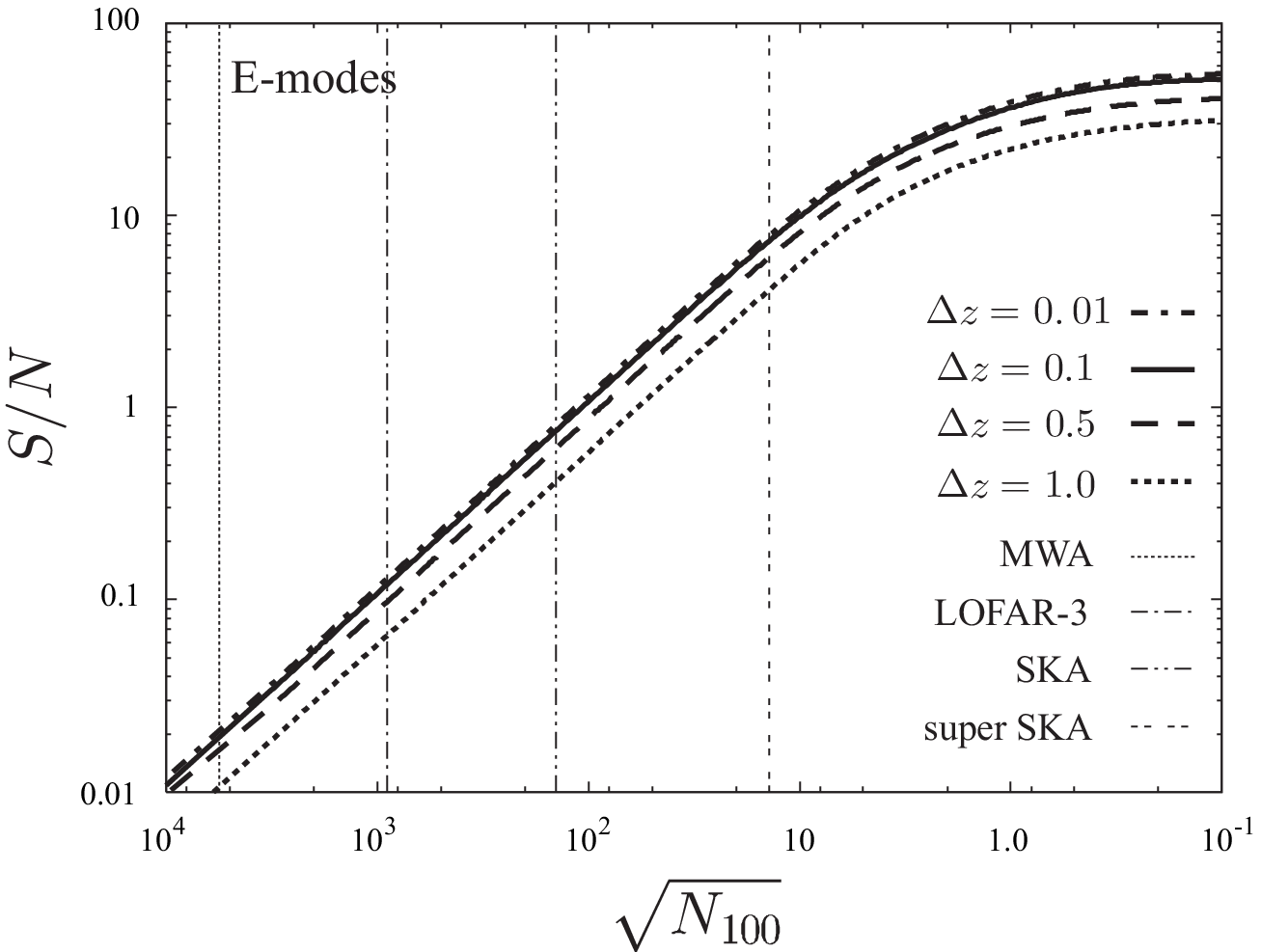}
  \end{center}
   \end{minipage}
  \end{tabular}
  \caption{The SN ratio as a function of $N_{100}$ for 
  the ideal full sky survey.
  We set $z_{\rm re} = 10$, $ f=0$ and $z_{\rm obs} = 10$. 
  The dashed-dotted, solid, dashed and dotted lines are for $ \Delta z= 0.01$, 
  $ \Delta z= 0.1$, $ \Delta z= 0.5$ 
  and $ \Delta z = 1$, respectively. The vertical lines represent the normalised noise power 
  spectra for each observation.
  The left panel shows the SN ratio for the 21-cm cross-correlation with the CMB Doppler anisotropy, 
  and the right panels is for the 21-cm cross-correlation with $E$-mode polarisation.}
  \label{fig:idealSN}
\end{figure}

In the estimation of the SN ratio of the cross-correlation, 
the auto-correlation for each observation is the
ultimate source of noise as shown in Eq.~\ref{eq:SNratio}. 
We therefore calculate the highest SN ratio
attainable, i.e. in the full sky survey (the sky
fraction is a multiplicative factor), and we plot the resulting SN
ratio as a function of $N_{100}$ in Fig.~\ref{fig:idealSN}. For this
computation, we set $z_{\rm obs}= 10$ and $z_{\rm re}= 10$ and $f=0$.
The amplitude of the 21-cm cross-correlation with the CMB Doppler
anisotropy depends on the reionisation duration.  Therefore, the
critical value of $N_{100}$, where the 21-cm auto-correlation-term ($C_\ell ^{21}$) 
dominates the 21-cm experimental noise ($N_\ell ^{21}$),
depends as well on the reionisation duration. The
critical value for $\Delta z =0.01$ is $N_{100} \sim 1.0$ and that for
$\Delta z =1.0$ is $N_{100} \sim 0.1$.  Since the 21-cm
cross-correlation with the CMB Doppler anisotropy has a sufficiently
high amplitude and a peak at large scales, it can be detected by
present or planned experiments (Fig.~\ref{fig:idealSN} left panel).
For the 21-cm cross-correlation with the CMB $E$-mode polarisation, 
although the long
duration of reionisation damps the power at high
$\ell$s, the noise which dominates the
cross-correlation signal at these scales makes it difficult to probe the duration. 
Therefore,
the difference due to the duration does not prominently appear
in the right panel of Fig.~\ref{fig:idealSN}.
The critical value of $N_{100}$ 
is same for different
reionisation durations (the critical value is $N_{100} \sim 1.0$).
Regardless of the duration of reionisation, 
the signal of the cross-correlation can be detected with an SN
ratio larger than 10. 

\section{conclusion}

We have investigated the detection level of the cross-correlation
between 21-cm fluctuations and large scale CMB anisotropy from the
EoR. 
We have evaluated the signal-to-noise (SN) ratio for the
  21~cm cross-correlation with both the Doppler temperature anisotropy
  and the $E$-mode polarisation. During the EoR, CMB anisotropies are
  also produced by patchy reionisation and Ostriker-Vishniac
  effect. These anisotropies also cross-correlate with 21 cm
  fluctuations, on small scales
  \citep{salvaterra-ciardi,cooray-2004,slosar-cooray-2007,jelic-zaroubi-2009}.
  However on such scales, the CMB anisotropy is contaminated by other
  secondary effects from galaxy clusters, e.g.~Sunyaev-Zeld'vich
  effect, which has a cross-correlation with 21-cm fluctuations
  \citep{slosar-cooray-2007}.  The detection of the cross-correlation
  signal from EoR at small scales is beyond the scope of the present
  study.  We will address this issue in a forthcoming paper.

For the cross-correlation between the 21-cm fluctuations and the CMB
Doppler anisotropy produced during the EoR, the amplitude of the
spectrum depends on the reionisation duration.  Short durations imply
high amplitude of the cross-correlation, and consequently large SN
ratio.  The cross-correlation between Planck and LOFAR, in its present
configuration, is sensitive to an ``instantaneous'' reionisation (with
$\Delta z =0.01$) only.  If the instrumental noise of LOFAR were
reduced by a factor ten, LOFAR could detect the cross-correlation
signal from the instantaneous reionisation with $S/N \sim 3$ for
single observation field and $S/N \sim 5$ for multi observation field.
Moreover, an ideal experiment with a sensitivity 10 times better and a
field of view twice as big as that of SKA can detect the signal from
the reionisation with $\Delta z = 0.5$.

For the cross-correlation between the 21-cm fluctuations and the CMB
$E$-mode polarisation, the angular power spectrum is damped on small
scales by the reionisation duration.  On those scales, the noise from
the primordial CMB polarisation dominates the cross-correlation signal
and makes the measurement of the cross-correlation insensitive to the
reionisation duration.  However, instead of the measurement of the
damping, the signal detection over several frequencies by an ideal
experiment 10 times more sensitive than SKA may give constraints on
the reionisation duration.

\end{document}